\documentclass{mn2e}
\usepackage{graphics}

\begin{document}

\title[{\em Spitzer} Imaging of $z\approx 6$ galaxies]{{\em Spitzer} Imaging of $i'$-drop 
Galaxies: Old Stars at $z \approx 6$}

\author[Eyles et al.\ ]{Laurence P.\ Eyles$^{1}$,
Andrew J.\ Bunker\,$^{1}$, Elizabeth R.\ Stanway\,$^{2}$,
Mark Lacy\,$^{3}$, \\
\newauthor
Richard S.\ Ellis\,$^{4}$, Michelle Doherty\,$^{5}$\\
$^{1}$\,School of Physics, University of Exeter, Stocker Road, Exeter, EX4 4QL,
U.K.\\ {\tt email:
eyles@astro.ex.ac.uk}\\
$^{2}$\,Astronomy Department, University of Wisconsin-Madison, 475 N.\ Charter Street, Madison, WI~53706, U.S.A. \\
$^{3}$\,Spitzer Science Center,California Institute of Technology, Mail Code 220-6, 1200 E.\ California Blvd., Pasadena, CA~91125, U.S.A.\\
$^{4}$\,California Institute of
Technology, Mail Stop 169-327, Pasadena, CA~91109, U.S.A. \\
$^{5}$\,Institute of Astronomy, Madingley Road, Cambridge, CB3 0HA, U.K.
}

\date{Accepted for publication in MNRAS, 15 July 2005}

\maketitle

\begin{abstract}
  We present new evidence for mature stellar populations with ages
  $>100$\,Myr in massive galaxies ($M_{\rm
    stellar}>10^{10}\,M_{\odot}$) seen at a time when the Universe was
  less than 1\,Gyr old. We analyse the prominent detections of two
  $z\approx 6$ star-forming galaxies (SBM03\#1\,\&\,\#3) made at
  wavelengths corresponding to the rest-frame optical using the IRAC
  camera onboard the {\em Spitzer Space Telescope}.  We had previously
  identified these galaxies in {\em HST}/ACS GOODS images of {\em Chandra}
  Deep Field South through the ``$i'$-drop'' Lyman break technique,
  and subsequently confirmed spectroscopically with the Keck
  telescope.  The new {\em Spitzer} photometry reveals significant
  Balmer/4000\,\AA\ discontinuities, indicative of dominant stellar
  populations with ages $>100$\,Myr.  Fitting a range of population
  synthesis models (for normal initial mass functions) to
  the {\em HST/Spitzer} photometry yields ages of
  $250-650$\,Myr and implied formation redshifts $z_{f}\approx
  7.5-13.5$ in presently-accepted world models.  Remarkably, our
  sources have best-fit stellar masses of
  $1.3-3.8\,\times10^{10}\,M_{\odot}$ (95\% confidence) assuming a
  Salpeter IMF.  This indicates that at least some
  galaxies with stellar masses $>20$\% of those of a present-day $L^*$
  galaxy had already assembled within the first Gyr after the Big
  Bang. We also deduce that the past average star formation rate must
  be comparable to the current observed rate (${\rm SFR}_{\rm UV}\sim
  5-30\,M_{\odot}\,{\rm yr}^{-1}$), suggesting that there may have
  been more vigorous episodes of star formation in such systems at
  higher redshifts. Although a small sample, limited primarily by
  {\em Spitzer's} detection efficiency, our result lends support to the
  hypothesis advocated in our earlier analyses of the Ultra Deep Field
  and GOODS {\em HST}/ACS data.  The presence of established systems
  at $z\approx 6$ suggests
  long-lived sources at earlier epochs ($z>7$) played a key role in
  reionizing the Universe.
\end{abstract}
\begin{keywords}

galaxies: evolution --
galaxies: formation --
galaxies: starburst --
galaxies: individual: SBM03\#1, SBM03\#3, GLARE\#3001, GLARE\#3011  --
galaxies: high redshift --
galaxies: stellar content
\end{keywords}

\section{Introduction}
\label{sec:intro}

In recent years, advances in detector efficiency, large ground-based
telescopes and space-based observatories such as the {\em Hubble Space
  Telescope (HST)\ } and {\em Spitzer Space Telescope}, have
revolutionized studies of the high redshift Universe. Searches based
on Lyman-$\alpha$ emission are now at last uncovering many galaxies at
$2<z<7$ (e.g., Kodaira et al.\ 2003; Rhoads et al.\ 2004). The
Lyman-break technique (Steidel, Pettini \& Hamilton 1995; Steidel et
al.\ 1996; 1999) has likewise proved successful in selecting
high-redshift star-forming galaxies. This utilises the rest-UV
continuum break seen shortwards of Lyman-$\alpha$ and caused by
H{\scriptsize~I} absorption in the intergalactic medium.
The redshift range $z\approx 6$ is of great importance, as it heralds
the end of the reionization of the Universe (Becker et al.\ 2001;
Kogut et al.\ 2003) which might be achieved through star formation.
Using {\em HST} and the new Advanced Camera for Surveys (ACS; Ford et
al.\ 2003), the Lyman-break technique has been pushed to this early
epoch (Stanway, Bunker \& McMahon 2003; Bouwens et al.\ 2004a; Yan \&
Windhorst 2004; Giavalisco et al.\ 2004) by using the $i'$ and $z'$
filters to isolate $i'$-drop galaxies.  Ground-based follow-up
spectroscopy (Bunker et al.\ 2003; Stanway et al.\ 2004a,b; Dickinson
et al.\ 2004) has shown that a colour cut of $(i'-z')_{AB}>1.5$\,mag
reliably finds star-forming galaxies at $z\approx 6$ with modest
foreground contamination (primarily from low-mass stars, and passively
evolving galaxies at $z\sim 1-2$). Subsequent low-dispersion slitless 
spectroscopy with  {\em HST}/ACS  (Malhotra et al.\ 2005) confirms the 
nature of $i'$-drops as  $z\sim 6$ galaxies with a small contaminant 
fraction of stars and low-redshift sources.

Yet an important part of the puzzle is missing: the $i'$-band drops are
selected in the rest-frame UV, and are therefore known to be actively
forming stars.  However, it is unclear whether these objects suffer
from significant reddening due to dust (in which case the star
formation rates will have been underestimated), or if there is an
underlying older stellar population which has been recently
rejuvenated. Publicly-available {\em Spitzer} imaging with the
Infrared Array Camera (IRAC; Fazio et al.\ 2004) as part of the Great
Observatories Origins Deep Survey\footnote{see {\tt
    http://www.stsci.edu/ftp/science/goods/}} (GOODS; Dickinson \&
Giavalisco 2003; Dickinson et al.\ {\em in prep}) allows us to address
both questions.

A key benefit of {\em Spitzer} photometry arises because the IRAC camera
samples wavelengths longwards of the age-sensitive Balmer \& 4000\AA\ 
breaks at $z\approx 6$. Accordingly, we have analysed the IRAC images
of GOODS-South field (centred on the {\em Chandra} Deep Field South,
Giacconi et al.\ 2002), where we have already selected $i'$-drop
galaxies from the GOODS-South ACS images (Stanway, Bunker \& McMahon 2003)
and the Ultra-Deep Field (UDF; Bunker et al.\ 2004).

Our goal in this paper is thus to focus on the infrared properties of
four $z\approx 6$ galaxies for which there are robust spectroscopic
redshifts, based on Lyman-$\alpha$ emission. The properties of the
entire $i'$-drop population in the GOODS fields is considered in a
forthcoming paper (Eyles et al.\ {\em in prep}). Our four confirmed
sources include two sources from Stanway, Bunker \& McMahon (2003):
the brightest confirmed $i'$-drop in the GOODS-South field (with
$z'=24.7$\,mag) SBM03\#3 (with a spectroscopic redshift of $z=5.78$
from Keck/DEIMOS, Bunker et al.\ 2003); and the brightest $i'$-drop in
the UDF, SBM03\#1 with $z'=25.3$\,mag (spectroscopically confirmed 
with Keck/DEIMOS by Stanway et al.\ 2004a as having a redshift of $z=5.83$). 
Both of these spectroscopic redshifts were independently confirmed by 
Dickinson et al.\ (2004). The other two sources come from the Gemini 
Lyman-$\alpha$ at Reionization Era (GLARE, Stanway et al.\ 2004b) 
spectroscopy with Gemini/GMOS: GLARE\#3001 ($z=5.79$, $z'=26.4$\,mag) 
and GLARE\#3011 ($z=5.94$, $z'=27.2$\,mag).

A plan of the paper follows. In Section~\ref{sec:reduct} we describe
the {\em Spitzer} imaging data and the methods used to fit stellar
populations to the broad band photometry derived collectively from
{\em Spitzer}, {\em HST} ACS \& NICMOS images, and ground-based
near-infrared data. We discuss the implications of the age and stellar
mass estimates in Section~\ref{sec:discuss}. Our conclusions are
presented in Section~\ref{sec:conclusions}. Throughout we adopt the
standard ``concordance'' cosmology\footnote{Adopting the first-year
  WMAP best-fit cosmology from Spergel et al.\ (2003) with
  $\Omega_M=0.27$, $\Omega_{\Lambda}=0.73$, and $H_0\,=\,71\,{\rm
    km\,s^{-1}\,Mpc^{-1}}$ makes negligible difference (the masses and
  luminosities would be 4 per cent greater, and the Universe would be
  4 per cent younger at $z=5.8$).} of $\Omega_M=0.3$,
$\Omega_{\Lambda}=0.7$, and use $H_0\,=\,70\,{\rm
  km\,s^{-1}\,Mpc^{-1}}$ -- in this cosmology, the Universe today is
13.67\,Gyr old, and at $z=5.8$ the age was 992\,Myr. All magnitudes
are on the $AB$ system (Oke \& Gunn 1983).

\section{Observations, Data Reduction and Stellar Population Fitting}
\label{sec:reduct}

\subsection{{\em Spitzer} Observations}

We concentrate here on the shortest-wavelength {\em Spitzer} images of
GOODS-South taken with IRAC as part of the ``Super Deep'' Legacy programme
(PID 194, Dickinson et al.\ {\em in prep}). We have analysed the first
half of the GOODS-South {\em Spitzer} data, taken with 39 Astronomical
Observation Requests (AORs) observed between 8--16 February 2004.

The IRAC camera comprises four channels, each with a $256^2$ InSb
array imaging a $5.2'\times 5.2'$ field with a pixel size of $\approx
1\farcs22$. Images were taken through four broad-band infrared
filters, with central wavelengths at approximately $\lambda_{\rm
  cent}=3.6\,\mu$m, $4.5\,\mu$m, $5.6\,\mu$m and $8.0\,\mu$m (channels
1--4), and widths of $\Delta\lambda_{\rm
  FWHM}=0.68,0.87,1.25,2.53\,\mu$m respectively.  The individual frame
times for each channel were 200\,s (except those taken at 8.0\,$\mu$m,
which comprise 4 integrations of 50\,s at each position).  Over the
course of the 39 AORs, a large $2\times 2$ mosaic of pointings was
executed, with smaller random sub-dithers when the pattern was
repeated, giving a $10'\times 10'$ coverage for each filter. Each AOR
comprised 10--11 pointings. Channels 1 \& 3 (3.6\,$\mu$m \&
5.6\,$\mu$m) have the same pointing, offset by $6.7'$ from the common
pointing of channels 2 \& 4 (4.5\,$\mu$m \& 8.0\,$\mu$m). As a result,
only a portion of the field is observed in all four wavebands.  Once
data from the second epoch (when the telescope has rotated by 180
degrees) has been gathered, images of the entire $10'\times 16.5'$
field matching the GOODS-South ACS survey (Giavalisco 2003) in all four
filters will be available.  Meanwhile, by design, the overlap region
common to all filters covers the Ultra Deep Field (Beckwith et al.\ 
2003).  The total exposure time in each channel is
$\approx$\,86\,ksec, depending on location.

\subsection{Data Reduction}

For our investigation, we used the pipeline-processed IRAC images at the
`Post-Basic Calibrated Data' (PBCD) stage\footnote{The PBCD images are 
available from 
\newline \tt http://ssc.spitzer.caltech.edu/popular/goods/irac$\_$04/}, 
details of which can be
found in the Infrared Array Camera Data Handbook, Version 1.0 (Reach
et al.\ 2004). The main steps in the pipeline include dark current
subtraction, application of flatfields, flux calibration (in units of
MJy/sr) and mosaicing of the individual frames of each AOR after
application of a distortion correction.  PBCD sets have a refined
pointing solution to an accuracy of $0\farcs2$ derived from 2MASS
point source catalogue objects in the field of view. Outlier rejection
is performed during the mosaicing process.

We used IRAF to combine the 39 PBCD mosaics (one for each AOR), which
were taken at different telescope roll angles. The IRAF task {\tt
  wregister} was used to rotate the frames to a common roll angle.
The individual registered frames were scaled to the same integration
time and all the frames were offset by the median counts in the centre
of the area (which was common to all frames) to account for floating
bias levels, and the 39 frames were combined using {\tt imcombine}
(weighting by the exposure times). During this, residual unrejected
cosmic rays, bad pixels, and the column pull-down and muxbleed
detector effects were removed using a percentile clipping method
``pclip'', rejecting at the $3\sigma$ level. The PBCD images were
converted from units of surface brightness (MJy/sr) into flux units of
$\mu$Jy per pixel by multiplying the data units by ${10^{12}}\times$
(pixel solid angle): a numerical factor of 34.98.  The images were
matched to the v1.0 reduced $z'$-band tiles of the GOODS-South field
released by the GOODS team\footnote{Available from \newline \tt
  ftp://archive.stsci.edu/pub/hlsp/goods/v1}. The astrometry was found
to be consistent to within $\approx 0\farcs2$. While our reduction was
underway, the GOODS team released an enhanced dataset
(DR1)\footnote{For the enhanced dataset GOODS DR1 see \newline {\tt
    http://data.spitzer.caltech.edu/popular/goods/ \newline
    20050209\_enhanced\_v1/Documents/goods\_dr1.html}.}, employing a
`multidrizzle' technique similar to that used successfully on {\em
  HST}/ACS GOODS data. This provides combined images with a pixel
scale of $\approx 0\farcs6$. The magnitudes listed in this paper are
determined from this ``drizzled'' data, and we have used our
independent reduction as a consistency check. Upon cross-checking the
astrometry, photometry and noise properties, they were found to be
consistent to 0.1\,mag (for bright sources) and $0\farcs2$.

In the final GOODS-South co-added DR1 `drizzled' images, we measured the 
FWHM of the PSF to be $\approx 1\farcs 5$ in channels 1 \& 2
(3.6\,\&\,4.5\,$\mu$m), and $\approx 1\farcs 8$ in channel 3
(5.6\,$\mu$m) \& $\approx 2\farcs 1$ in channel 4 (8\,$\mu$m).  Even
for the short-wavelength IRAC images, the {\em Spitzer} PSF is much
larger than the typical size of the $z\approx 6$ $i'$-band drop
galaxies (whose half-light radii $r_{hl}<0\farcs2$, Bunker et al.\ 
2004). At {\em Spitzer} resolution, these galaxies are clearly
unresolved and so we treat them as point sources. The galaxy images in
the various wavebands are shown in Figures~\ref{fig:SBM03_1stamp},
\ref{fig:SBM03_3stamp} \& \ref{fig:GLARE3001stamp}.

\begin{figure}
\resizebox{0.48\textwidth}{!}{\includegraphics{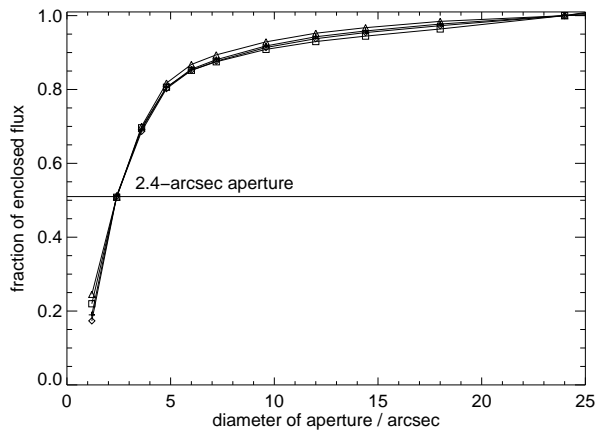}}
\caption{The fractional flux enclosed within a circular aperture
  as a function of aperture diameter, determined from 4 bright but
  unsaturated isolated point sources in the IRAC channel 1
  (3.6\,$\mu$m) GOODS-South image. The small aperture used for faint
  sources ($2\farcs4$-diameter) is marked: this encircles 52 per cent
  of the energy, implying an aperture correction of 0.7\,mags. Note
  that the official {\em Spitzer} calibration for IRAC uses a
  $24''$-diameter aperture, so the fraction of the flux there is set
  to 1.0 by definition.}
\label{fig:radprof}
\end{figure}

\begin{figure}
\resizebox{0.48\textwidth}{!}{\includegraphics{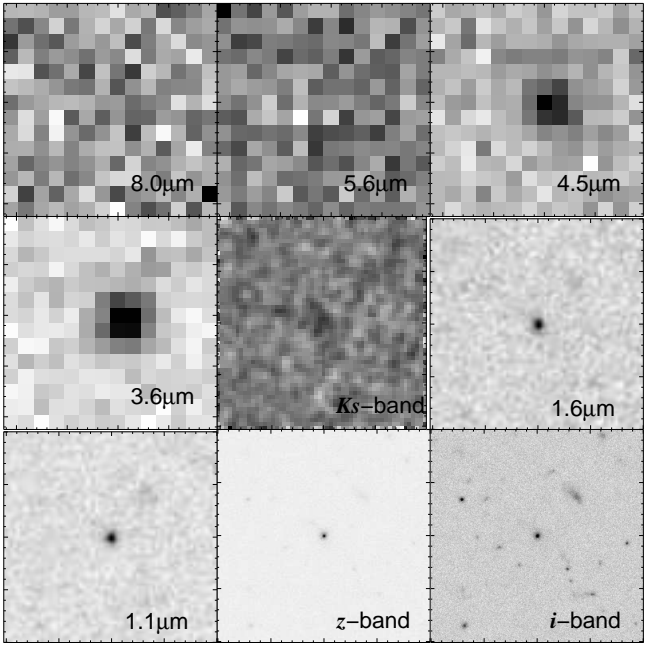}}
\caption{Images of SBM03\#1 ($z=5.83$), taken with {\em HST}/ACS ($i'$- and
 $z'$-band); {\em HST}/NICMOS (1.1\,$\mu$m F110W `$J$-band', and 1.6\,$\mu$m 
F160W `$H$-band'),
 VLT/ISAAC $K_s$-band (smoothed with a 3-pixel $0\farcs45$ boxcar) and 
the 4 {\em Spitzer} channels (3.6--8.0\,$\mu$m). Each panel is 8\,arcsec 
across (a projected distance of 50\,kpc).}
\label{fig:SBM03_1stamp}
\end{figure}

\begin{figure}
  \resizebox{0.48\textwidth}{!}{\includegraphics{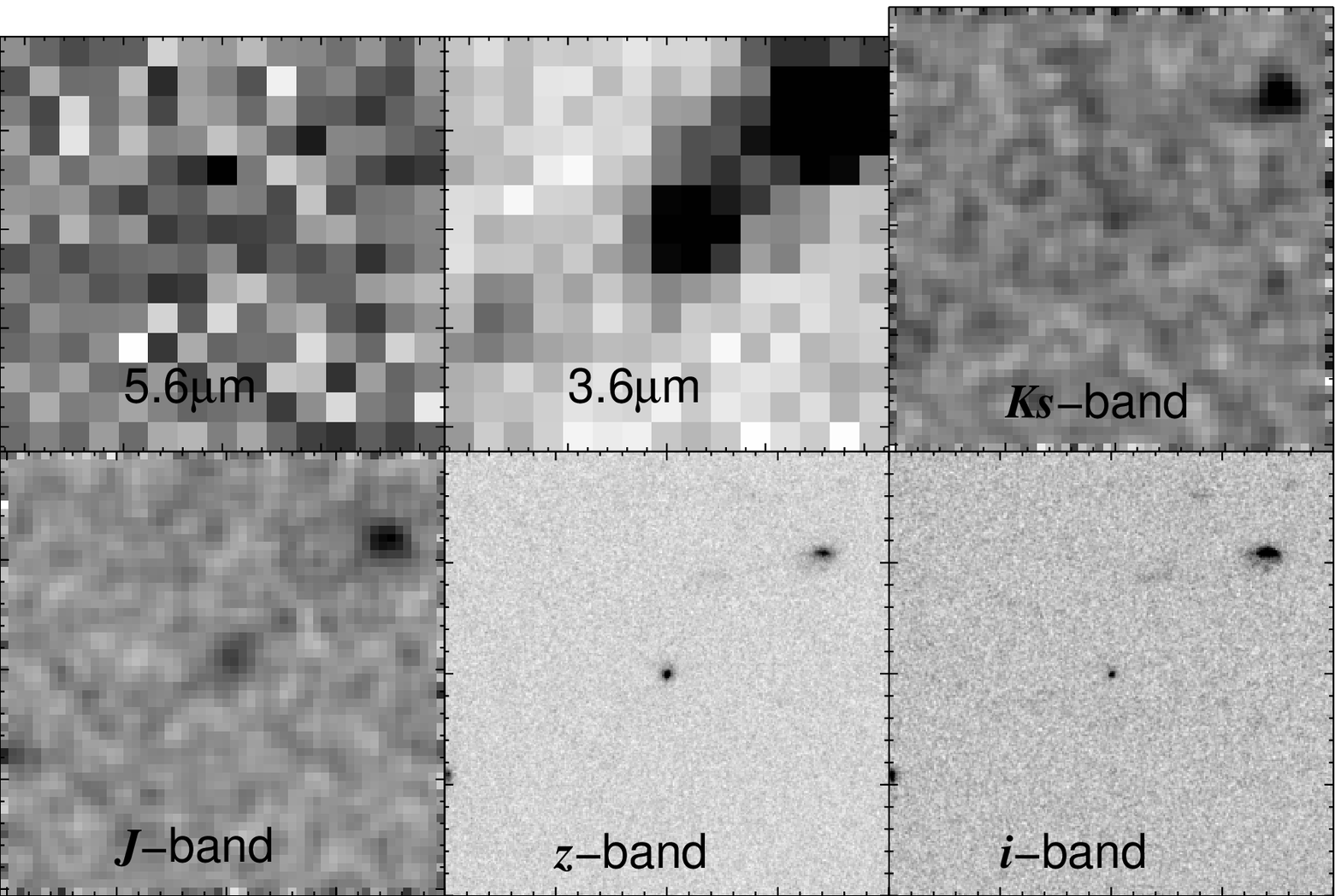}}
\caption{Images of SBM03\#3 ($z=5.78$), taken with {\em HST}/ACS ($i'$- and
 $z'$-band); VLT/ISAAC $J$- and $K_s$-bands and {\em Spitzer} channels 
3.6\,\&\,5.6\,$\mu$m. Each panel is 8\,arcsec across (a projected distance 
of 50\,kpc).  The $J$ and $K_{s}$ band images have been smoothed using a 
3-pixel  box.}
\label{fig:SBM03_3stamp}
\end{figure}

To construct spectral energy distributions of our 4
spectroscopically-confirmed $i'$-band drop galaxies in the GOODS-South
field (detailed in Section~\ref{sec:intro} and
Table~\ref{tab:targets}) we undertook aperture photometry in the
various images. In order to maximize the signal-to-noise ratio ($S/N$)
and minimize possible confusion with other foreground objects, we used
a diameter $\approx 1.5\times {\rm FWHM}$ for the IRAC images,
appropriate for unresolved objects. The aperture diameters were 4, 4,
5 \& 6 `drizzled' pixels for the 4 channels (3.6, 4.5, 5.6 \&
8.0\,$\mu$m), corresponding to $2\farcs4$, $2\farcs 4$, $3\farcs 0$,
\& $3\farcs 7$ respectively.  We used the IRAF {\tt digiphot.phot} package to
measure the enclosed flux at the exact coordinates determined from the
GOODSv1.0 and UDF $z'$-band images, taking the residual background
from an annulus between $12''$ and $24''$ radius. We applied aperture
corrections to compensate for the flux falling outside the aperture:
these were $\approx 0.7$\,mag, as determined from bright but
unsaturated point sources in the images, with apertures of diameter
$18''$ in channels 1 \& 2, $24''$ in channel 3, and $30''$ in channel
4. The curve of growth for four stars is shown in
Figure~\ref{fig:radprof} (note that the official {\em Spitzer} calibration
for IRAC also uses a similar $24''$-diameter aperture, so the aperture
correction there is 1.0 by definition).  These aperture corrections
are consistent with those derived for the First Look Survey (Lacy et
al.\ 2005).

\begin{table*}
\begin{tabular}{l|c|c|c|c|c|}
ID & RA (J2000) & Dec (J2000) & Redshift & Age of Universe / Gyr \\
\hline\hline
SBM03\#1$^{\dagger}$ & 03:32:40.01 & -27:48:15.0 & 5.83 & 0.95 \\
SBM03\#3 & 03:32:25.61 & -27:55:48.7 & 5.78 & 0.96 \\
GLARE\#3001 & 03:32:46.04 & -27:49:29.7 & 5.79 & 0.96 \\
GLARE\#3011 & 03:32:43.10 & -27:45:17.6 & 5.94 & 0.93 \\
\end{tabular}

$^{\dagger}$\,this is also known as GLARE\#1042 (Stanway et al.\ 2004b),
\#20104 (Bunker et al.\ 2004), SiD002 (Dickinson et al.\ 2004) and YW01a 
(Yan \& Windhorst 2004).
\caption{The four spectroscopically-confirmed $i'$-drop galaxies in 
the GOODS-South Field.}
\label{tab:targets}
\end{table*}

\begin{figure}
  \resizebox{0.48\textwidth}{!}{\rotatebox{90}{\includegraphics*[54,54][287,738]{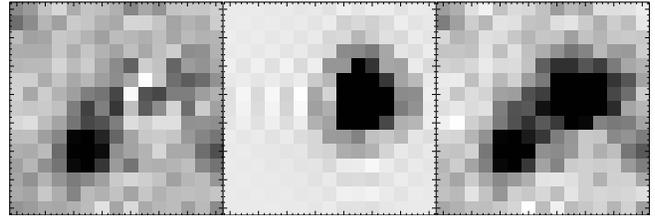}}}
\caption{{\em Spitzer} channel 1 (3.6\,$\mu$m) image of SBM03\#3 (right, lower-left 
  of panel), with the best-fit GALFIT model for the surface brightness
  profile of the nearby neighbour (middle image, an exponential disk
  convolved with the PSF), and the {\em Spitzer} image with the neighbouring
  galaxy subtracted (left). The photometry of SBM03\#3 is only
  affected at the $\sim$\,5\,per cent level by confusion from the
  neighbour.  The frame is 9\,arcsec across.}
\label{fig:SBM03_3_subtract}
\end{figure}

\begin{figure}
  \resizebox{0.48\textwidth}{!}{\includegraphics{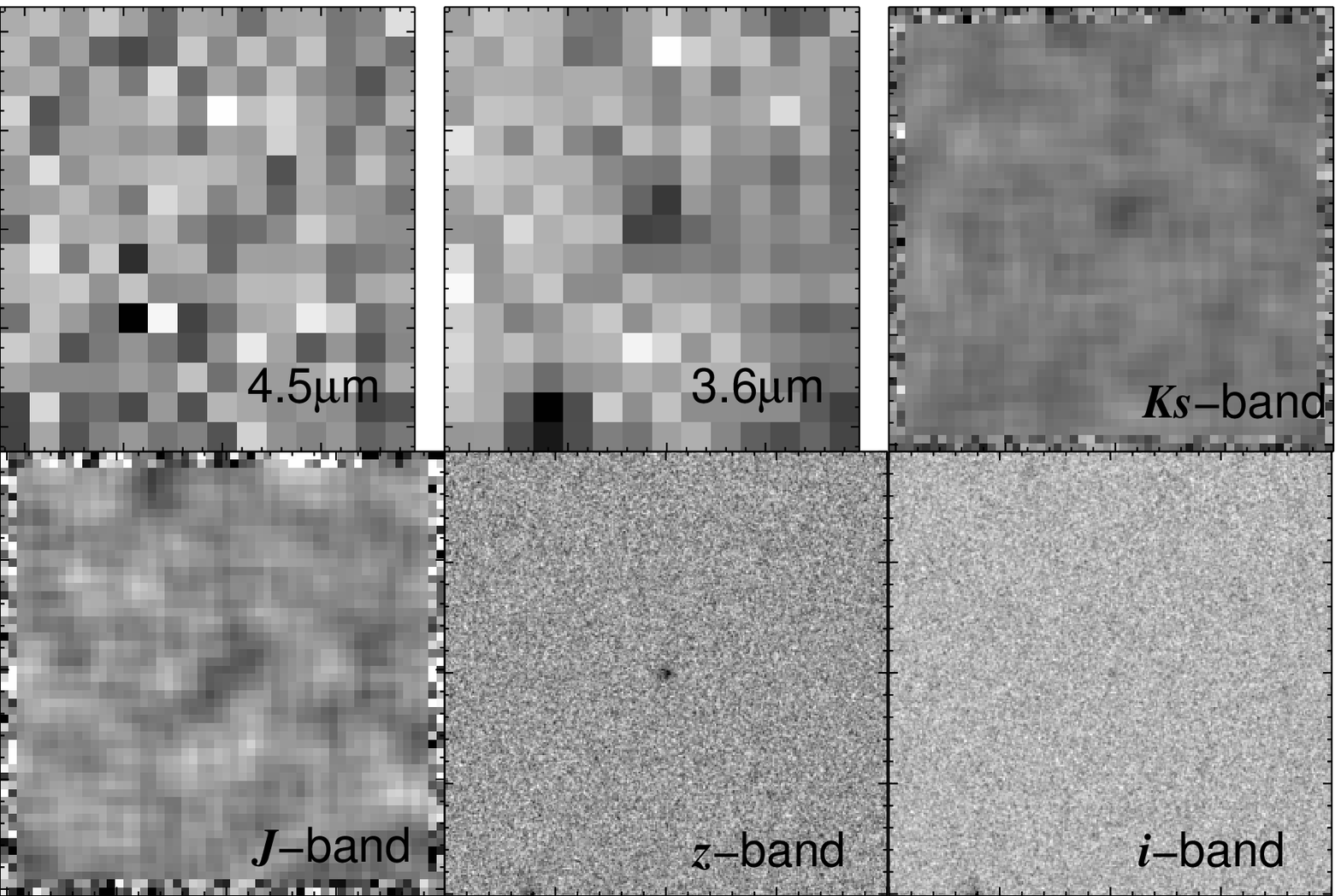}}
\caption{Images of GLARE\#3001 ($z=5.79$), taken with {\em HST}/ACS ($i'$- 
and $z'$-band); VLT/ISAAC $J$- and $K_s$-bands and {\em Spitzer} channels 
3.6\,\&\,4.5\,$\mu$m. Each panel is 8\,arcsec across (a projected distance 
of 50\,kpc). The $J$ and $K_{s}$ band images have been smoothed using a 
5-pixel box.}
\label{fig:GLARE3001stamp}
\end{figure}

Checks were made to ensure the objects were not contaminated by the
presence of any neighbouring bright foreground sources. An object was
deemed to be uncontaminated if no bright source lay within a $4''$
radius from it. All four objects satisfied this criteria; the bright
source close to SBM03\#3 lies on the $4''$ limit
(Figure\,\ref{fig:SBM03_3stamp}).  The {\em Spitzer} magnitudes in
Table\,\ref{tab:photometry} have the neighbour subtracted out (i.e.,
are deblended) but this has only a $\approx 5$\% effect on the
measured flux in our $2\farcs4$-diameter aperture; less than the error
bar. The subtraction of the neighbour was done using the ``GALFIT''
software (Peng et al.\ 2002). This takes into account the PSF (we
give it a PSF for the drizzled image derived from stacking 4 bright
but isolated stars) and a best-fit galaxy surface brightness profile. 
An exponential disk worked best in the case of this neighbour, but it
is not very sensitive to the exact functional form with the poor
resolution of {\em Spitzer}. Figure~\ref{fig:SBM03_3_subtract} shows the
postage-stamp image of SBM03\#3 with the model fit of the neighbour
subtracted.

The noise for each of the four channels was checked in two different
ways.  First, we derived an estimate based on a Poisson model using
the detector gain, number of frames combined, and the background
counts (adding back the zodiacal background estimate subtracted by the
pipeline but recorded in the header). Secondly, we measured the
standard deviation in background counts of the images.  As the
mosaicing process introduces correlations between pixels, we also made
noise estimates using the individual BCDs and assuming it decreased as
the square root of the number of frames. These estimates lead to
$3\sigma$ limiting AB magnitudes of 26.5 and 26.1 in $2\farcs 4$
apertures in channels 1 and 2, respectively, and 23.8 and 23.5 in
$3\farcs 0$ apertures in channels 3 and 4, respectively.  There will
be additional background fluctuations caused by faint galaxies (i.e.\ 
confusion noise), which will increase the noise. Both methods produce
consistent estimates.

Table~\ref{tab:photometry} lists the measured AB magnitudes (corrected
to approximate total magnitudes through an aperture correction) for
the IRAC $3.6\,\mu$m, $4.5\,\mu$m and $5.6\,\mu$m channels.  None of
the sources were detected at $>3\,\sigma$ in the 5.6\,\&\,8\,$\mu$m
channels, although SBM03\#3 has a very marginal ($2\,\sigma$)
detection at 5.6\,$\mu$m ($AB=24.4\pm 0.7$). In the short-wavelength
channels, SBM03\#1\,\&\,\#3 are well-detected ($>10\,\sigma$,
$AB\approx 24$), and we have a more marginal $3\,\sigma$ detection of
GLARE\#3001 ($z'_{AB}=26.4$) at $AB\approx 26$. Only SBM03\#1 is
detected at 4.5\,$\mu$m -- SBM03\#3 falls outside the region surveyed
so far in this filter, and GLARE\#3001 is undetected. The fainter
GLARE\#3011 ($z'_{AB}=27.2$) was undetected at all IRAC wavelengths.
For the {\em HST} photometry, we use the ACS $i'$-band and $z'$-band
magnitudes from Stanway, Bunker \& McMahon (2003) from GOODS-South and, in
the case of SBM03\#1, from the deeper UDF (Bunker et al.\ 2004).
These magnitudes are already corrected to a total flux, also through an
aperture correction.  We use NICMOS magnitudes (Thompson et al.\ 2005) in 
F110W and F160W (`$J$' and `$H$' band) from Stanway,
McMahon \& Bunker (2005) for SBM03\#1, and for the others we used v1.0
of the ESO VLT/ISAAC GOODS/EIS images\footnote{Available from \newline {\tt
    http://www.eso.org/science/goods/releases/20040430/}} in the $J$
and $K_{s}$ bands (Vandame et al.\ {\em in prep}).  In the
ground-based near-infrared ISAAC images, we used $1''$-diameter
apertures. The seeing varied across the ISAAC field, as different
tiles were taken over many nights, so we determined the aperture
correction from unresolved sources in each tile. For the $J$- and
$K_s$-band images the seeing is typically good
(${\rm FWHM}=0\farcs4-0\farcs5$), and the aperture correction is $\approx
0.3-0.5$\,mag, determined from bright but unsaturated isolated
 stars measured in 6$''$-diameter apertures.

The galaxy SBM03\#3 is towards the edge of the released ESO/VLT
imaging, where fewer frames overlap and the noise is higher. Its $K_s$
magnitude seems to be anomalously faint (a $3\,\sigma$ detection at
$K_{AB}=25.5$, twice as faint as the shorter-wavelength $z'$ and $J$,
which have $AB=24.7$\,mag, and four times fainter than the
longer-wavelength IRAC 3.6\,$\mu$m detection at $AB=24.0$). We still
include the $K_s$ filter in the fitting of the stellar populations
(Section~\ref{sec:SEDs}), but caution that the $K_s$ magnitude may not
be reliable. Fortunately, this filter makes minimal difference to the
best-fit populations as its statistical weighting is low.

\subsection{Spectral Energy Distributions}
\label{sec:SEDs}

The final step in the reduction process is the construction of
spectral energy distributions (SEDs) for the chosen sources. Spitzer
photometry and SED fitting has already produce great insight into the
$z=2-3$ Lyman-break star-forming galaxies (Barmby et al.\ 2004;
Shapley et al.\ 2005) and here we apply similar techniques at
$z\approx 6$. In order to compare our photometry with stellar spectral
synthesis models, we utilise the latest Bruzual \& Charlot (2003,
hereafter B\&C) isochrone synthesis code.  We use the Padova
evolutionary tracks (preferred by B\&C 2003).  The models utilise 221
age steps from $10^5$ to $2\times 10^{10}$\,yr, approximately
logarithmically spaced. Models with Salpeter (1955) initial mass
functions (IMF) were selected, although in Section~\ref{sec:dust} we
also consider the effect of adopting a Chabrier (2003) IMF. There are
6900 wavelength steps, with high resolution (FWHM 3\,\AA ) and 1\,\AA\ 
pixels over the wavelength range of 3300\,\AA\ to 9500\,\AA\ (unevenly
spaced outside this range).  From the full range of metallicities
offered by the code, we considered both solar and 1/5th solar models.
From several star formation histories available, a single stellar
population (SSP -- an instantaneous burst), a constant star formation
rate (SFR), and exponentially decaying ($\tau$) SFR models were used.
This latest version of the Bruzual \& Charlot (2003) models includes
an observationally-motivated prescription for thermally-pulsing
asymptotic giant branch (TP-AGB) stars (Liu, Graham \& Charlot 2002),
and also the TP-AGB multi-metallicity models of Vassiliadis \& Wood
(1993).

For each of the four $i'$-drops with spectroscopic redshifts, the
filters were corrected to their rest-frame wavelengths by the
appropriate redshift factor.  The measured flux was folded through the
filter transmission profiles, and the best-fit age model was computed
by minimising the reduced $\chi^{2}$, using the measured errors on the
magnitudes. The number of degrees of freedom is the number of
independent data points (magnitudes in different wavebands). The
Bruzual \& Charlot spectra are normalized to an initial mass of
$1\,M_{\odot}$ for the instantaneous burst SSP model, and an SFR of
$1\,M_{\odot}\,{\rm yr}^{-1}$ for the continuous star formation model.
The fitting routine returned the normalisation for the model which was
the best-fit to the broad band photometry (i.e., minimized the reduced
$\chi^{2}$) -- this normalisation was then used to calculate the
corresponding best-fit stellar mass (see Section~\ref{sec:mass}).

For SBM03\#1, the SED-fitting process was conducted using photometry 
from the detections in the $z'$, F110W, F160W, $K_{s}$, 3.6\,$\mu$m 
\& 4.5\,$\mu$m filters, whilst for SBM03\#3 the $z'$, $J$, $Ks$, 
3.6\,$\mu$m and 5.6\,$\mu$m data were used.

Although some of our data points (particularly from the {\em HST}/ACS
imaging) have $S/N>10$, we set the minimum magnitude error to be
$\Delta({\rm mag}) = 0.1$ to account for flux calibration uncertainties.
During the SED-fitting process, the $i'$-band flux was ignored, as this band is
prone to contamination due to Lyman-$\alpha$ forest absorption
shortwards of Lyman-$\alpha$ ($\lambda_{\rm rest}=1216$\,\AA) and also
emission line contamination due to Lyman-$\alpha$ itself.

\begin{table*}
\centering
\begin{tabular}{l|c|c|c|c|c|c|c|c|c|}
ID & $i'$ & $z'$ & $J$ & $K_{s}$ & 3.6\,$\mu$m & 4.5\,$\mu$m & 5.6\,$\mu$m\\
\hline\hline
SBM03\#1 & $26.99\pm 0.03$ & $25.35\pm 0.02$ & $25.54^{\dagger}\pm 0.04$ & $25.14\pm 0.22$ & $24.24\pm 0.09$ & $24.42\pm 0.15$ &  $>23.8$ ($3\,\sigma$) \\
SBM03\#3 & $26.27\pm 0.13$ & $24.67\pm 0.03$ & $24.72\pm 0.15$ & $25.55\pm 0.40^{\diamond}$ & $23.94\pm 0.07$ & --- & $24.39\pm 0.69$ \\
GLARE\#3001 & $28.03\pm 0.14$ & $26.37\pm 0.06$ & $26.11\pm 0.32$ & $25.22\pm 0.32$  & $25.88\pm 0.36$ &  $>26.1$ ($3\,\sigma$) & $>23.8$ ($3\,\sigma$) \\
GLARE\#3011 & $>28.8$ ($3\,\sigma$) & $27.15\pm 0.12$ & $>26.2$ ($3\,\sigma$) & $>25.6$ ($3\,\sigma$) & $>26.5$ ($3\,\sigma$) & $>26.1$ ($3\,\sigma$) & $>23.8$ ($3\,\sigma$) \\
\end{tabular}

$^{\diamond}$\,The $K_s$ magnitude of SBM03\#3 is anomalously faint.\newline
$^{\dagger}$\,From NICMOS F110W imaging (Stanway, McMahon \& 
Bunker 2004), rather than ESO VLT/ISAAC imaging. The NICMOS F160W magnitude of SBM03\#1 is $H_{AB}=25.51\pm 0.05$
\caption{Magnitudes (AB system) of four $i'$-band drop galaxies. SBM03\#3 is 
not in the  field of view for the 4.5\,$\mu$m and 8.0\,$\mu$m filters. All four galaxies are undetected at 8.0\,$\mu$m, with AB $>$ 23.5 ($3\,\sigma$).} 
\label{tab:photometry}
\end{table*}

\begin{table*}
\centering
\begin{tabular}{l|c|c|c|c|c|c|c|}
Model & Reduced $\chi^{2}_{min}$ & Age / Myr & E(B-V) & $M_{\rm tot}$ / $10^{10}M_{\odot}$ & $M_{\rm stellar}$ / $10^{10}M_{\odot}$ & Current SFR / $M_{\odot}\,{\rm yr}^{-1}$ & $b$ / $10^{-10} {\rm yr}^{-1}$ \\
\hline\hline
burst (SSP) & 2.83 & 90.5 & 0.00 & 1.35 & 1.13 & 0.00$^{\dagger}$ & 0.00 \\
$\tau=10$\,Myr & 2.81 & 102 & 0.00 & 1.34 & 1.14 & 0.05$^{\dagger}$ & 0.04 \\
$\tau=30$\,Myr & 2.34 & 143 & 0.02 & 1.63 & 1.36 & 4.59 & 3.38 \\
$\tau=70$\,Myr & 1.61 & 255 & 0.00 & 2.05 & 1.68 & 7.88 & 4.69 \\
$\tau=100$\,Myr & 1.38 & 321 & 0.00 & 2.33 & 1.97 & 9.80 & 4.20 \\
$\tau=300$\,Myr & 0.94 & 641 & 0.00 & 3.41 & 2.70 & 15.2 & 4.97 \\
$\tau=500$\,Myr & 0.90 & 905$^{\diamond}$ & 0.00 & 4.35 & 3.38 & 17.0 & 5.03 \\
$\tau=1000$\,Myr & 0.97 & 1280$^{\diamond}$ & 0.00 & 5.46 & 4.21 & 21.1 & 5.01 \\
continuous & 1.15 & 5000$^{\diamond}$ & 0.00 & 7.93 & 5.84 & 15.9 & 2.72 \\
\end{tabular}

$^{\dagger}$ Ruled out because SFR below lower limit set by Lyman-$\alpha$ emission.\\
$^{\diamond}$ Ruled out as age is close to or exceeds 1\,Gyr, age of Universe at $z\approx 6$
\caption{Favoured model parameters of various SED fits for SBM03\#1, 
metallicity $=Z_{\odot}$. We varied age between 0 and 20 Gyr, and 
extinction between 0.00 and 1.00. We list both the mass currently in stars 
($M_{\rm stellar}$), and the total 
baryonic mass ($M_{\rm total}$ which includes the mass returned to the 
IGM by evolved stars)
for each star formation history.}
\label{tab:SBM03_1_solar}
\end{table*}

\begin{table*}
\centering
\begin{tabular}{l|c|c|c|c|c|c|c|}
Model & Reduced $\chi^{2}_{min}$ & Age / Myr & E(B-V) & $M_{\rm tot}$ / $10^{10}M_{\odot}$ & $M_{\rm stellar}$ / $10^{10}M_{\odot}$ & Current SFR / $M_{\odot}\,{\rm yr}^{-1}$ & $b$ / $10^{-10} {\rm yr}^{-1}$ \\
\hline\hline
burst (SSP) & 1.54 & 114 & 0.00 & 1.32 & 1.09 & 0.00$^{\dagger}$ & 0.00 \\
$\tau=10$\,Myr & 1.50 & 128 & 0.00 & 1.35 & 1.13 & 0.02$^{\dagger}$ & 0.02 \\
$\tau=30$\,Myr & 1.30 & 161 & 0.01 & 1.42 & 1.17 & 2.23$^{\dagger}$ & 1.91 \\
$\tau=70$\,Myr & 0.99 & 255 & 0.00 & 1.59 & 1.29 & 6.12 & 4.74 \\
$\tau=100$\,Myr & 0.87 & 321 & 0.00 & 1.83 & 1.47 & 7.70 & 5.24 \\
$\tau=300$\,Myr & 0.77 & 641 & 0.00 & 2.77 & 2.18 & 12.4 & 5.69 \\
$\tau=500$\,Myr & 0.78 & 905$^{\diamond}$ & 0.00 & 3.57 & 2.76 & 14.0 & 5.07 \\
$\tau=1000$\,Myr & 0.81 & 1280$^{\diamond}$ & 0.00 & 4.54 & 3.48 & 17.5 & 5.03 \\
continuous & 0.93 & 2600$^{\diamond}$ & 0.02 & 4.34 & 3.35 & 16.7 & 4.99 \\
\end{tabular}

$^{\dagger}$ Ruled out because SFR below lower limit set by Lyman-$\alpha$ emission.\\
$^{\diamond}$ Ruled out as age is close to or exceeds 1\,Gyr, age of Universe at $z\approx 6$
\caption{Favoured model parameters of various SED fits for SBM03\#1, 
metallicity $=0.2 Z_{\odot}$. We varied age between 0 and 20 Gyr, and 
extinction between 0.00 and 1.00.We list both the mass currently in stars 
($M_{\rm stellar}$), and the total 
baryonic mass ($M_{\rm total}$ which includes the mass returned to the 
IGM by evolved stars)
for each star formation history.}
\label{tab:SBM03_1_subsolar}
\end{table*}

\begin{table*}
\centering
\begin{tabular}{l|c|c|c|c|c|c|c|}
Model & Reduced $\chi^{2}_{min}$ & Age / Myr & E(B-V) & $M_{\rm tot}$ / $10^{10}M_{\odot}$ & $M_{\rm stellar}$ / $10^{10}M_{\odot}$  & Current SFR / $M_{\odot}\,{\rm yr}^{-1}$ & $b$ / $10^{-10} {\rm yr}^{-1}$ \\
\hline\hline
burst & 3.66 & 47.5 & 0.00 & 1.12 & 0.97 & 0.00$^{\dagger}$ & 0.00 \\
$\tau=10$\,Myr & 3.65 & 64.1 & 0.00 & 1.17 & 1.02 & 1.93$^{\dagger}$ & 1.89 \\
$\tau=30$\,Myr & 3.21 & 114 & 0.00 & 1.53 & 1.29 & 11.7 & 9.07 \\
$\tau=70$\,Myr & 2.75 & 203 & 0.00 & 2.22 & 1.84 & 18.6 & 10.1 \\
$\tau=100$\,Myr & 2.55 & 255 & 0.00 & 2.54 & 2.08 & 21.5 & 10.3 \\
$\tau=300$\,Myr & 2.06 & 453 & 0.00 & 3.54 & 2.85 & 33.4 & 11.7 \\
$\tau=500$\,Myr & 1.96 & 641 & 0.00 & 4.78 & 3.77 & 36.8 & 9.76 \\
$\tau=1000$\,Myr & 1.88 & 905$^{\diamond}$ & 0.00 & 6.81 & 5.37 & 46.3 & 8.62 \\
continuous & 1.82 & 1800$^{\diamond}$ & 0.00 & 5.20 & 4.02 & 28.9 & 7.19 \\
\end{tabular}

$^{\dagger}$ Ruled out because SFR below lower limit set by Lyman-$\alpha$ emission.\\
$^{\diamond}$ Ruled out as age is close to or exceeds 1\,Gyr, age of Universe at $z\approx 6$
\caption{Favoured model parameters of various SED fits for SBM03\#3, 
metallicity $=Z_{\odot}$. We varied age between 0 and 20 Gyr, and 
extinction between 0.00 and 1.00. We list both the mass currently in stars 
($M_{\rm stellar}$), and the total 
baryonic mass ($M_{\rm total}$ which includes the mass returned to the 
IGM by evolved stars)
for each star formation history.}
\label{tab:SBM03_3_solar}
\end{table*}

\begin{table*}
\centering
\begin{tabular}{l|c|c|c|c|c|c|c|}
Model & Reduced $\chi^{2}_{min}$ & Age / Myr & E(B-V) & $M_{\rm tot}$ / $10^{10}M_{\odot}$ & $M_{\rm stellar}$ / $10^{10}M_{\odot}$ & Current SFR / $M_{\odot}\,{\rm yr}^{-1}$ & $b$ / $10^{-10} {\rm yr}^{-1}$ \\
\hline\hline
burst & 2.15 & 71.9 & 0.00 & 1.38 & 1.17 & 0.00$^{\dagger}$ & 0.00 \\
$\tau=10$\,Myr & 2.12 & 80.6 & 0.00 & 1.30 & 1.12 & 0.41$^{\dagger}$ & 0.37 \\
$\tau=30$\,Myr & 1.92 & 128 & 0.00 & 1.51 & 1.26 & 7.21 & 5.72 \\
$\tau=70$\,Myr & 1.73 & 203 & 0.00 & 1.81 & 1.50 & 15.2 & 10.1 \\
$\tau=100$\,Myr & 1.67 & 255 & 0.00 & 2.07 & 1.69 & 17.5 & 10.4 \\
$\tau=300$\,Myr & 1.50 & 453 & 0.00 & 2.96 & 2.37 & 27.9 & 11.8 \\
$\tau=500$\,Myr & 1.46 & 571 & 0.00 & 3.65 & 2.90 & 34.2 & 11.8 \\
$\tau=1000$\,Myr & 1.41 & 806$^{\diamond}$ & 0.00 & 5.37 & 4.22 & 43.3 & 10.3 \\
continuous & 1.36 & 1400$^{\diamond}$ & 0.00 & 4.34 & 3.44 & 25.1 & 7.30 \\
\end{tabular}

$^{\dagger}$ Ruled out because SFR below lower limit set by Lyman-$\alpha$ 
emission.\\
$^{\diamond}$ Ruled out as age is close to or exceeds 1\,Gyr, age of 
Universe at $z\approx 6$
\caption{Favoured model parameters of various SED fits for SBM03\#3, 
metallicity $=0.2 Z_{\odot}$. We varied age between 0 and 20 Gyr, and 
extinction between 0.00 and 1.00. We list both the mass currently in stars 
($M_{\rm stellar}$), and the total 
baryonic mass ($M_{\rm total}$ which includes the mass returned to the 
IGM by evolved stars)
for each star formation history.}
\label{tab:SBM03_3_subsolar}
\end{table*}

\begin{figure}
\resizebox{0.48\textwidth}{!}{\includegraphics{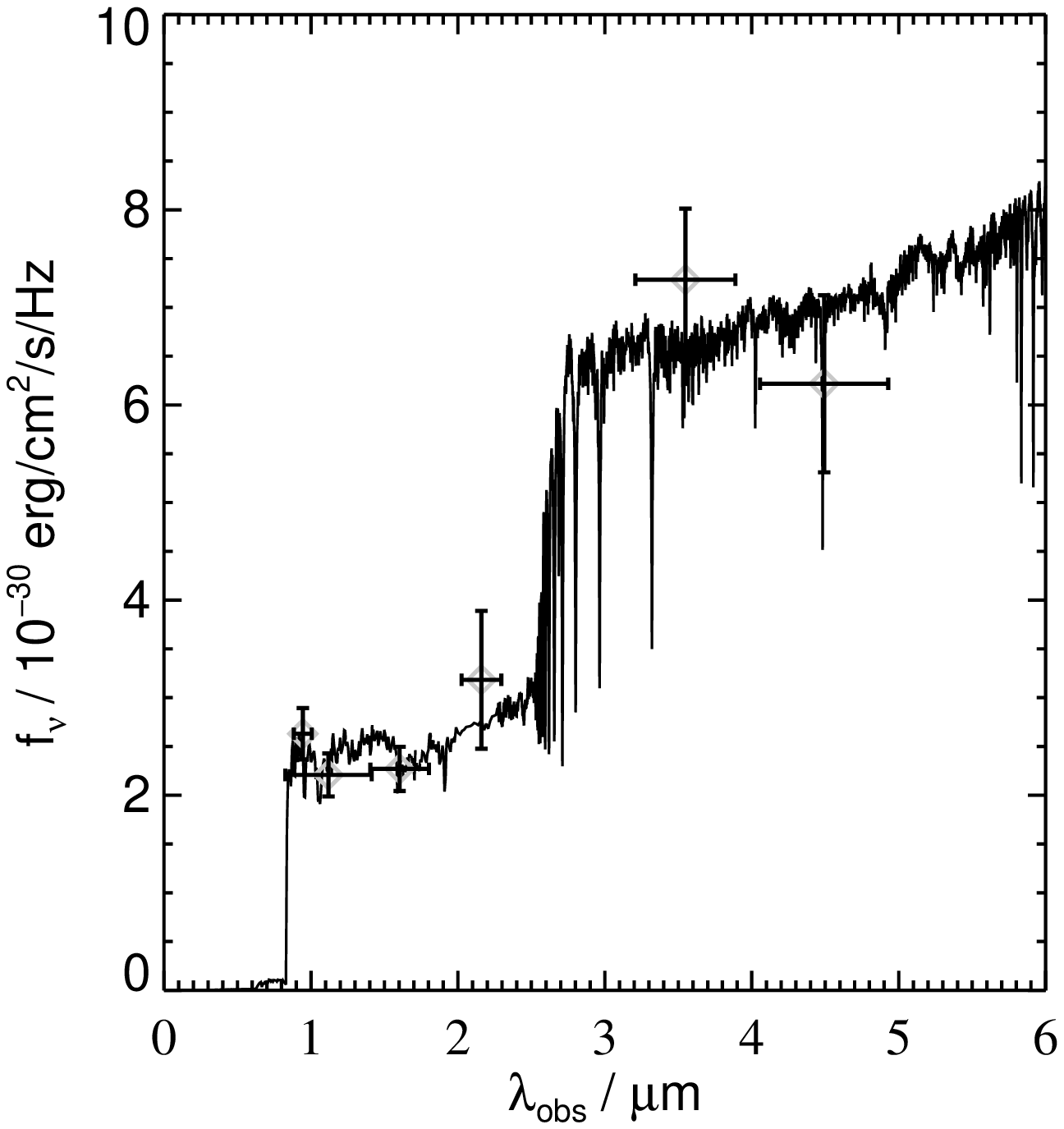}}
\caption{Best-fit Bruzual \& Charlot model for SBM03\#1: an exponentially 
decaying star formation rate with $\tau = 300$\,Myr, viewed 640\,Myr after
 the onset of 
star formation. The stellar mass is $3.4\times 10^{10}\,M_{\odot}$.
Flux density is in $f_{\nu}$ units.}
\label{fig:flnuSBM03_1}
\end{figure}

\begin{figure}
\resizebox{0.48\textwidth}{!}{\includegraphics{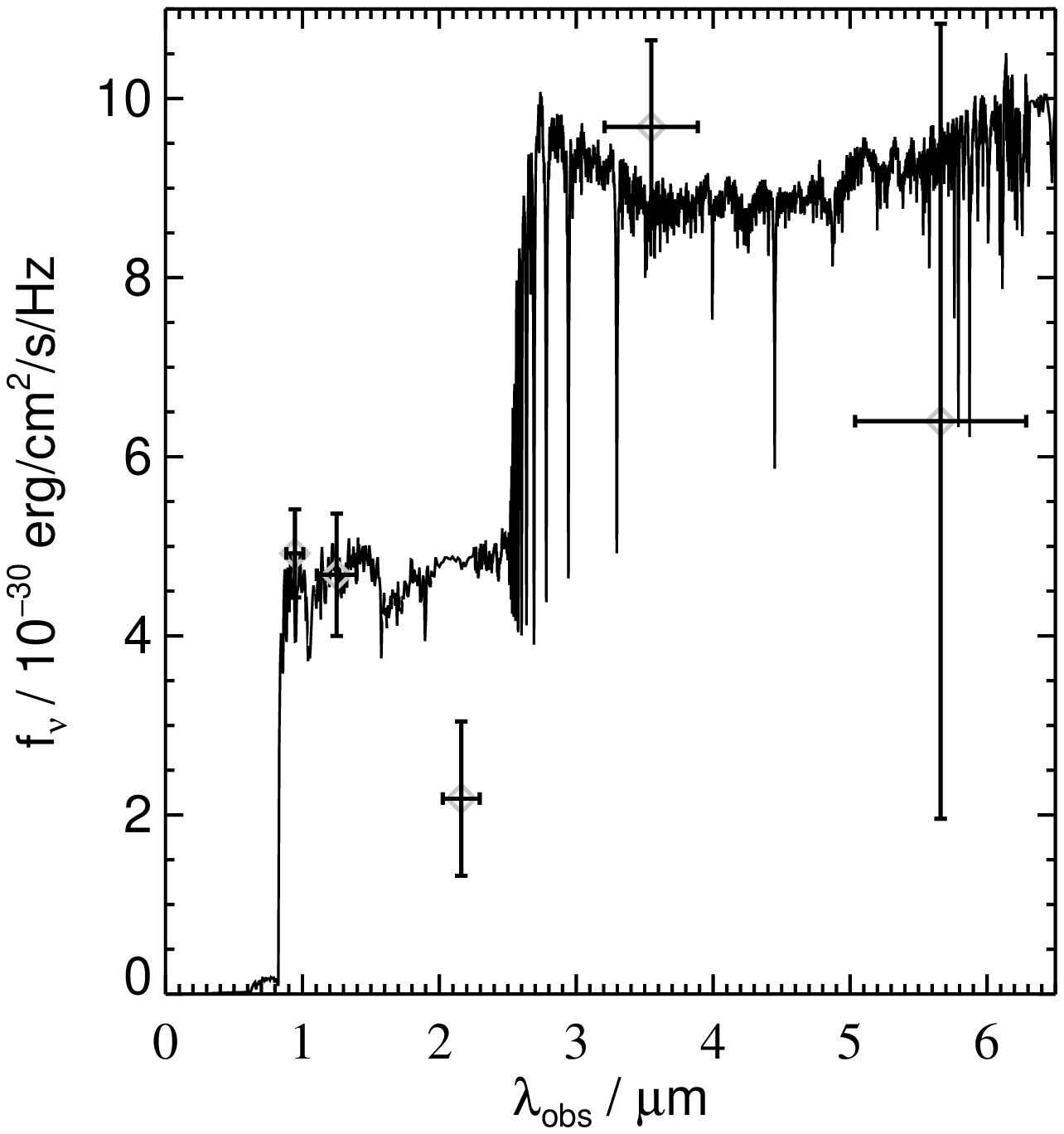}}
\caption{Best-fit Bruzual \& Charlot model for SBM03\#3: an exponentially
 decaying star formation rate with $\tau = 500$\,Myr, viewed 640\,Myr after 
the onset of star formation, and with a stellar mass of 
$4.8\times 10^{10}\,M_{\odot}$. Flux density is in $f_{\nu}$ units.}
\label{fig:flnuSBM03_3}
\end{figure}

\begin{figure}
\resizebox{0.48\textwidth}{!}{\includegraphics{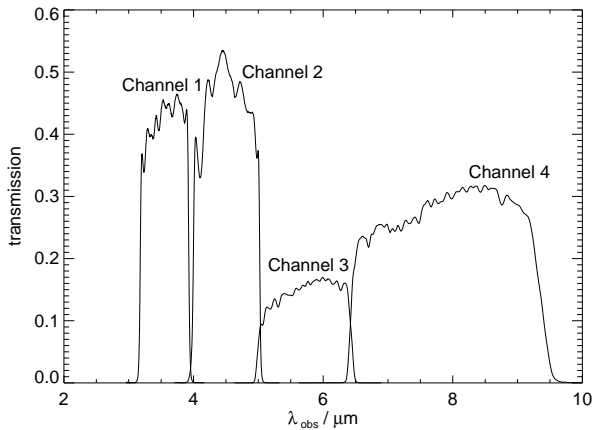}}
\caption{Filter profiles of the four {\em Spitzer} infrared wavebands.}
\label{fig:filtprof}
\end{figure}

\begin{figure}
\resizebox{0.48\textwidth}{!}{\includegraphics{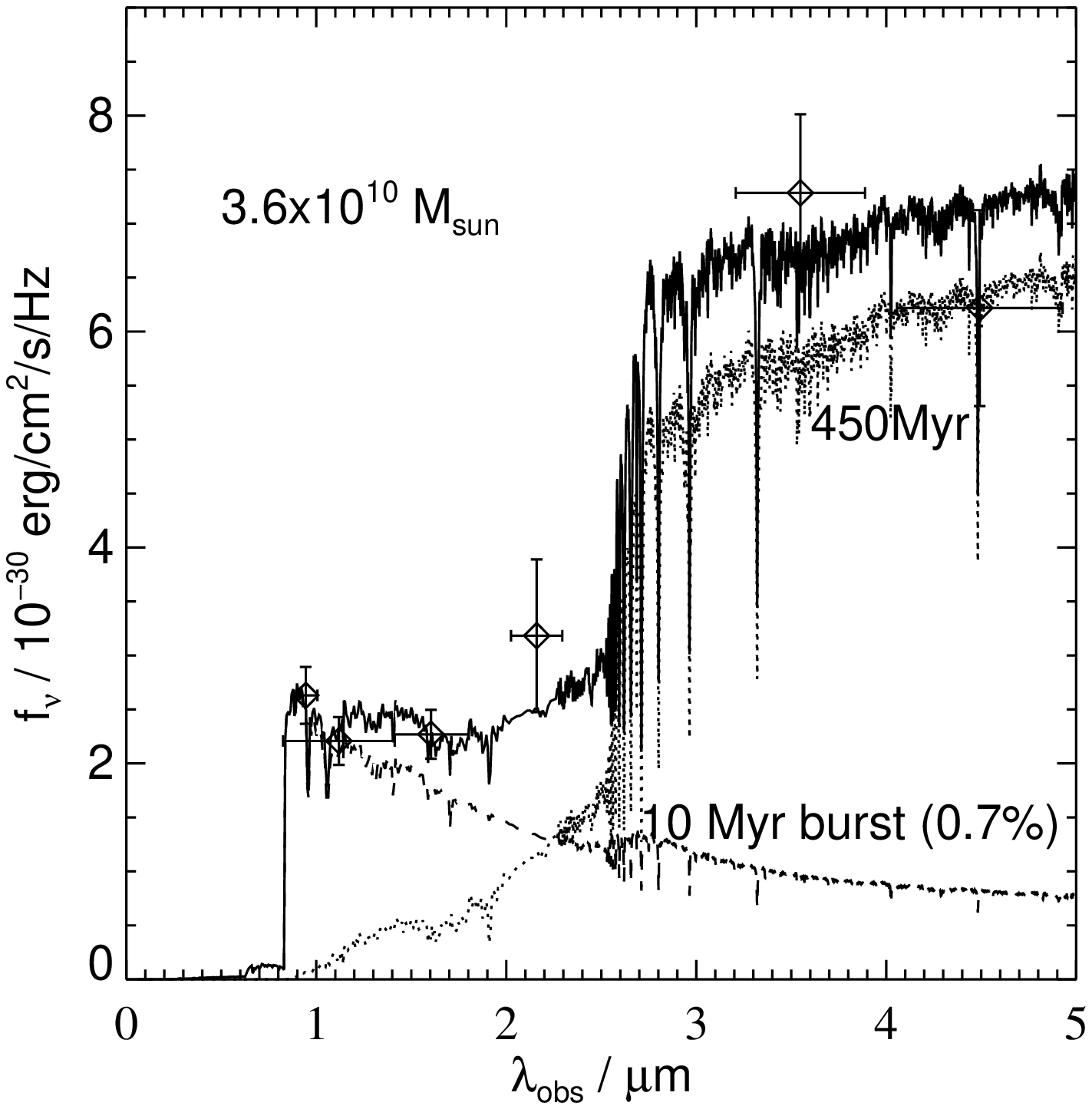}}
\caption{The best-fitting two-component stellar population model (Salpeter 
IMF) for SBM03\#1: a dominant 450\,Myr population of mass 
$3.6\times 10^{10}\,M_{\odot}$, with some ongoing star formation activity 
(a burst for the last 10\,Myr involving 0.7\% of the stellar mass). Using a 
Chabrier IMF produces an identical best-fit age, with a similar mass fraction 
in the burst (0.6\%), but a 30\% lower total stellar mass 
($2.5\times 10^{10}\,M_{\odot}$).}
\label{fig:twopopSBM03_1}
\end{figure}

\section{Analysis}
\label{sec:discuss}

For two of the spectroscopically-confirmed galaxies
(SBM03\#1\,\&\,\#3) we have robust {\em Spitzer} detections at
3.6\,$\mu$m of $AB\approx 24$\,mag ($>10\,\sigma$), with the fainter
GLARE sources marginally detected (GLARE\#3001) or undetected
(GLARE\#3011). Progressing in increasing wavelength from the {\em
  HST}/ACS $z'$-band and near-infrared ($0.9-2.2\,\mu$m) to the {\em
  Spitzer}/IRAC 3.6\,$\mu$m channel, the two well-detected $i'$-drops
brighten by $\Delta({\rm mag})_{AB}=0.8$
(Figures~\ref{fig:flnuSBM03_1}\,\&\,\ref{fig:flnuSBM03_3}) -- a factor
of $2$ in flux density, $f_{\nu}$.  SBM03\#1 is also robustly detected
with a similar magnitude at 4.5\,$\mu$m. We now consider the
implications of the spectral energy distributions of these $i'$-drops,
specifically which star formation histories can produce the observed
spectral breaks between $2.2-3.6\,\mu$m.

\subsection{Balmer/4000\,\AA\ Breaks in $z\approx 6$ Galaxies}
\label{sec:break}

The presence of a Balmer/4000\,\AA\ break is suggestive of a system
viewed a significant time after a major epoch of star formation -- the
Balmer break at 3648\,\AA\ is strongest in an A-star population, and
metal line blanketting (predominantly Fe{\scriptsize~II}) due to an
older late-type stellar population produces the 4000\,\AA\ break. For
the two $i'$-drops with the best {\em Spitzer} detections
(SBM03\#1\,\&\,\#3) we find evidence of a significant
Balmer/4000\,\AA\ break: the brightening by a factor of 2 in $f_{\nu}$
from the near-infrared ($\approx 0.9-2.2\,\mu$m) to 3.6\,$\mu$m implies a
break amplitude of $\approx 1.7$ (for $f_{\lambda}$ flux densities). The $z',\,J\,\&\,K_{s}$
colours are relatively flat in $f_{\nu}$
(Figures~\ref{fig:flnuSBM03_1}\,\&\,\ref{fig:flnuSBM03_3}) --ignoring
the discrepantly faint $K_s$ magnitude of SBM03\#3-- which favours the
spectral break interpretation rather than dust reddening (see
Section~\ref{sec:dust}). The break amplitude of $\approx 1.7$ is
comparable to that seen by Le Borgne et al.\ (2005) in massive
post-starburst galaxies at much lower redshifts ($z\sim 1$) from the
``Gemini Deep Deep Survey'' project. Indeed, the break amplitude in
our two significant $z\approx 6$ cases is only slightly less than the
$D4000$ (Bruzual 1983)
observed at $z\approx 0$ in the Sloan Digital Sky Survey (e.g., Kauffmann 
et al.\ 2003), with $D4000\approx 1.7-2.0$.

For all the galaxies discussed here, there has been recent or ongoing
formation of at least {\em some} massive stars prior to the epoch of
observation, as our Keck/Gemini spectra show Lyman-$\alpha$ emission produced
through photoionization of hydrogen by the short-lived OB stars
(Bunker et al.\ 2003; Stanway et al.\ 2004a). In order to produce
the Balmer/4000\,\AA\ break amplitude at 
$\lambda_{\rm rest}\approx 4000$\,\AA ,
most of the stellar mass probably formed well before the current
starburst, most likely $>100$\,Myr previously (Section~\ref{sec:age}).
Hence the galaxies SBM03\#1\,\&\,\#3 are already, to some extent,
established systems. This is a significant finding since, at $z\approx 6$,
the Universe is less than 1\,Gyr old.

There is no significant evidence for a Balmer/4000\,\AA\ break in the
two (fainter) spectroscopically-confirmed $i'$-drops: the
$3\,\sigma$ detection of GLARE\#3001 at $3.6\,\mu$m suggests a break
of $0\pm0.5$\,mag, and the non-detection of GLARE\#3011 places a
$3\,\sigma$ upper limit of $\Delta AB=1$\,mag on the break amplitude. Hence
the constraints for the stellar ages of GLARE\#3001\,\&\,3011 are
weak, but they are consistent with younger stellar populations than
SBM03\#1\,\&\,\#3, which are brighter and presumably more massive. In
the remainder of this discussion, we focus on these two brighter
sources which have significant {\em Spitzer} detections for which we can
estimate the stellar ages and masses.

\subsection{Ages of the $i'$-drop galaxies}
\label{sec:age}

We explored the best-fit stellar ages and star formation histories of
the galaxies SBM03\#1\,\&\,\#3 by comparing the SEDs from our
broad-band photometry with the population synthesis models of Bruzual
\& Charlot (2003).  We cover a range of star formation histories
(presented in
Tables\,\ref{tab:SBM03_1_solar},\,\ref{tab:SBM03_1_subsolar},\,\ref{tab:SBM03_3_solar},\,\ref{tab:SBM03_3_subsolar}\,\&\,\ref{tab:tau})
designed to bracket most plausible evolution scenarios. Only a subset
of these star formation histories provided acceptable fits to our
photometry. We now discuss several classes of model and their
validity.

We begin by considering an idealized ``simple stellar population''
(SSP) model, where a galaxy is viewed some time after its entire
stellar mass formed in an instantaneous burst. These models provide
relatively poor fits (reduced $\chi^2\approx 3$ for solar metallicity)
to our data. This is understandable given that we know that there is at least
{\em some} ongoing star formation in both galaxies. The SSP models provide
an absolute lower age limit for the bulk of the stellar mass. These
SSP models yield an age of $\approx 100$\,Myr for the best fit
population (and $>70$\,Myr at 99\% confidence) in the absence of dust
reddening (see Section~\ref{sec:dust}), implying a formation redshift
of $z_{f}\gg 6.4$.

At the other extreme, a constant star formation history was
considered. The best fit ages in this simplistic scenario are
1.5-5\,Gyr for metallicity $0.2-1\,Z_{\odot}$ (with SBM03\#3 best-fit
with a younger population than SBM03\#1, and lower metallicities
decreasing the age); these ages are obviously unphysical, as they
exceed the age of the universe at this redshift ($\approx 1$\,Gyr).
This strongly implies that we are seeing both galaxies at an epoch
when the star formation rate is declining, or that the current burst
was preceded by a more significant episode of star formation.

Hence we next considered a range of more realistic models including an
exponentially decaying star formation rate (SFR), such that the
current ${\rm SFR}_{t}={\rm SFR}_0\,e^{-t/\tau}$, where ${\rm SFR}_0$
is the star formation rate at the onset of the burst and $\tau$ is the
decay time in Myr. Such models are intermediate between SSP
($\tau\rightarrow 0$) and continuous star formation ($\tau\rightarrow
\infty$) models.  We considered several values of $\tau$ ranging
between 10\,Myr to 1\,Gyr. As before, those models with ages older
than the Universe at $z\approx 6$ were rejected.  Also, models with
$\tau = 10$\,Myr were disregarded as they yielded current SFRs below
the lower limit set by the Lyman-$\alpha$ emission
(Section~\ref{sec:SFRs}).  The favoured models have an age of $\approx
600$\,Myr, with decay timescales of $\tau = 300$\,Myr for SBM03\#1,
and $\tau = 500$\,Myr for SBM03\#3.

Finally, we consider a model where the galaxies are composed of two
distinct stellar components: an ongoing starburst at the time of
observation and an older population which formed in an instantaneous
burst some time previously.  We considered starbursts with a constant
SFR which started 3, 10, 30, \& 100\,Myr prior to the epoch of
observation.  We varied the ratio of total stellar masses in these two
populations, and found the best-fit ages for the old component to be
$400-500$\,Myr.  From the best-fit models (Table~\ref{tab:tau}), the
fraction of total stellar mass being formed in the starburst is
$0.5-5$\% for starbursts of $3-100$\,Myr duration.

In summary, our SED fitting shows that the broad-band colours can be
fit with a variety of stellar ages/star formation histories.  The
lower limit on the age is $>100$\,Myr (from an SSP model) with the
oldest allowed models comparable to the age of the Universe at
$z\approx 6$. Our best-fits to the broad-band photometry come from
exponentially-decaying star formation histories, or a two-component
model where only $0.5-5$\% of the stellar mass is forming in an
ongoing starburst.  These best-fit models have mean stellar ages of
$260-640$\,Myr, which would require formation redshifts of
$z_{f}\approx 7.5-13.5$.

It has been speculated that certain treatments of the
thermally-pulsing asymptotic giant branch (TP-AGB) may lead to
different age estimates from broad-band colours. Maraston (2005)
suggests the $z\sim 2-3$ population of `iEROs' identified by Yan et
al.\ (2005) in {\em HST+Spitzer} images can be fit by younger ages
than indicated by the Bruzual \& Charlot (2003) models. Potentially
the TP-AGB has a significant impact on the restframe $V-K$ colours
measured in the ACS/IRAC images at $z\approx 2-3$. However, our
$i'$-drops are at higher redshifts, and the TP-AGB has little effect
at $\lambda<0.7\,\mu$m (see fig.~14 of Maraston 2005), the wavelength
range which we are probing with our {\em HST+Spitzer} detections.

\begin{table*}
\centering
\begin{tabular}{l||c|c|c|c|c}
burst duration / $Myr$ & reduced $\chi^2$ & frac of mass & current SFR / $M_{\odot}$\,yr$^{-1}$ & age / $Myr$ & stellar mass / $10^{10}\,M_{\odot}$ \\
\hline\hline
3 & 0.71 & 0.5\% & 60.0 & 400 & 3.6 \\
10 & 0.68 & 0.7\% & 25.3 & 450 & 3.6 \\
30 & 0.72 & 1.7\% & 19.0 & 450 & 3.3 \\
100 & 0.79 & 5.2\% & 68.3 & 450 & 3.1 \\
\end{tabular}
\caption{Two-population composite models for SBM03\#1, assuming an ongoing 
burst of  constant star formation rate, commencing 3-100\,Myr ago, within an
older galaxy. The best-fit galaxy age and mass fraction of the starburst
for each model are tabulated (for solar metallicity, $Z_{\odot}$).}
\label{tab:tau}
\end{table*}

\subsection{Stellar Masses at $z\approx 6$}
\label{sec:mass}

\begin{figure}
\resizebox{0.48\textwidth}{!}{\includegraphics{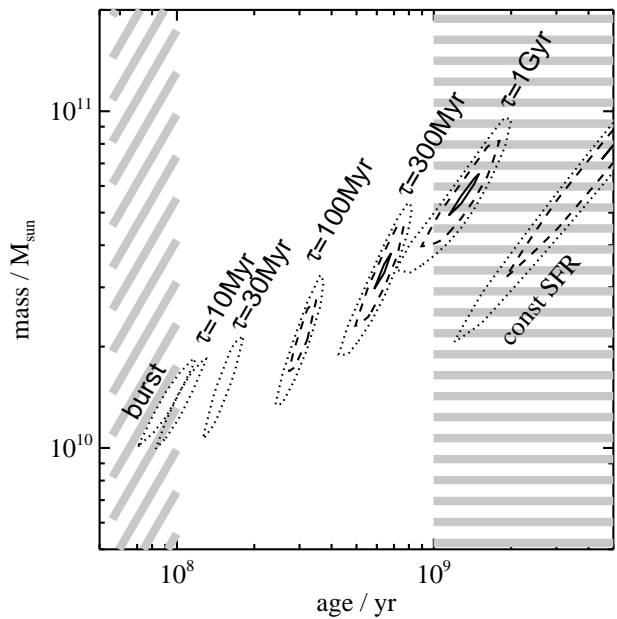}}
\caption{The allowed range of masses for several 
  exponentially-decaying SFR models for SBM03\#1, with decay times ranging from
  $\tau=10-1000$\,Myr, as well as an instantaneous burst model and a
  constant SFR model. Those with stellar ages $>1$\,Gyr (right shaded
  region) are ruled out (the Universe is younger than this at
  $z\approx 6$). The models with ages $<10^{8}$\,yr are also excluded
  as they fail to produce the observed current SFR inferred from
  Lyman-$\alpha$ emission (left shaded region). Contours are 68\%
  confidence (solid line), 95\% confidence (dashed line) and 99\%
  confidence (dotted line) for reduced $\chi^2$ of $\approx 1,2,3$.}
\label{fig:massplotSBM03_1}
\end{figure}

The SED fitting procedure described above leaves the normalisation of
the Bruzual \& Charlot model as a free parameter (along with the age
for a particular input model). Our code outputs this normalization
and, by using the luminosity distance to the $i'$-drop galaxies, we
can calculate the corresponding best fit stellar mass for each star
formation history. Tables~\ref{tab:SBM03_1_solar},
\ref{tab:SBM03_1_subsolar}, \ref{tab:SBM03_3_solar} \&
\ref{tab:SBM03_3_subsolar} give the best-fit masses returned by the
SED fitting.  We list both the mass currently in stars ($M_{\rm
  stellar}$, and the total baryonic mass ($M_{\rm total}$ which
includes the mass returned to the IGM by evolved stars) for each star
formation history. To assess the errors on the stellar mass estimates,
we took the best-fit mass, $M_{\rm stellar}$, and recalculated the
reduced $\chi^2$ for the same age, metallicity and star formation
history, but using total masses in the range $0.1-3\,M_{\rm stellar}$
(shown in Figure~\ref{fig:massplotSBM03_1}).  In all our models, the
stellar masses of SBM03\#1\,\&\,3 were $>10^{10}\,M_{\odot}$ at 95\%
confidence ($2\,\sigma$). The lowest masses were returned by the SSP
model, and the exponentially-decaying models with short decay times
($\tau\sim 10$\,Myr): in fact, these models are unable to provide
sufficient ongoing star formation to explain the lower limit set on
the star formation rate by the Lyman-$\alpha$ emission (see
Section~\ref{sec:SFRs}). Our preferred models ($\tau=70-500$\,Myr and
the two-component stellar population) have stellar masses of $\approx
2-4\times 10^{10}\,M_{\odot}$ for a Salpeter IMF (we consider the
impact of adopting a Chabrier IMF in Section~\ref{sec:dust}).

We measure stellar masses from the best-fitting SEDs of $M>2\times
10^{10}\,M_{\odot}$ at $z\approx 6$. This is surprisingly large,
supporting our contention that at least these two objects are
well-established galaxies. The stellar mass is equivalent to 20\% of
that for a $L^*$ galaxy today, using $L^{*}_{r} = -21.21$ from the
SDSS analysis of Blanton et al.\ (2003) and taking $M/L_{V}\approx
5\,M_{\odot}/L_{\odot}$ (appropriate for a $\approx 10$\,Gyr old
population from B\&C models using a Salpeter IMF) to obtain
$M^{*}=1.2\times 10^{11}\,M_{\odot}$, comparable to the estimate of
$M^{*}=1.4\times 10^{11}\,M_{\odot}$ from Cole et al.\ (2001) for our
adopted Salpeter IMF.

\subsection{Stellar Mass Density and the Kormendy Relation}
\label{sec:kormendy}

A key question given the increased information content on our two
$z\approx 6$ sources is the likely descendant population and
connection with systems seen at lower redshift. In this respect, we
now consider the high surface brightness and small angular extent of
the $i'$-drop galaxies.

To facilitate this discussion, we examine the properties of the
$z\approx 6$ galaxies with reference to the $z\sim 0$ Kormendy
relation between galaxy surface brightness and half-light radius
(Kormendy 1977). Both SBM03\#1\,\&\,\#3 have half-light radii
(effective radii) of $r_{hl}\approx 0\farcs08$ (Bunker et al.\ 2004;
Bunker et al.\ 2003), and IRAC total magnitudes of $AB\approx 24$ at
$3.6\,\mu$m (corresponding to the rest-frame $V$-band). The average
surface brightness {\em within} the effective radius, $\langle
I\rangle_{hl}$, is related to the average surface brightness {\em at}
$r_{hl}$, $I_{hl}$, by $\langle I_{hl}\rangle = 3.6\times \langle
I\rangle_{hl}$ for a de Vaucouleurs profile (Ciotti \& Bertin 1999).
Hence our surface brightness at $r_{hl}$ is $\mu_{e}=21.6\,{\rm
  mag\,arcsec}^{-2}$.

Cosmological surface brightness dimming of $(1+z)^4$ would have dimmed
{\em real} surface brightnesses by $+8.3$\,mag, and passive evolution
from an age of $\approx 200$\,Myr at $z=5.8$ to $12.8$\,Gyr at $z=0$
would produce a luminosity dimming of a factor of $\times 30$ in $V$
($+3.7$\,mag, using the B\&C models). Hence, we might expect
SBM03\#1\,\&\,\#3 to have surface brightnesses {\em today} of $\mu_{e}
= 21.6 -8.3 +3.7 = 17.0\,{\rm mag\,arcsec}^{-2}$ for their half-light
radii of $r_{hl}=0.4$\,kpc. In fact, the passively-evolved surface
brightnesses sit comfortably on the present-day Kormendy relation
(e.g., that of Ziegler et al.\ 1999 for cluster ellipticals),
extrapolated to smaller scale lengths. Hence it seems that the
inferred properties of these $z\approx 6$ galaxies are compatible with
galaxy scaling relations at $z=0$, subject to dimming through stellar
evolution.  Although our analysis has focused on only two of the most
luminous systems, chosen by virtue of their luminosity, the possible
implications are profound. What do these $i'$-drop galaxies at
$z\approx 6$ evolve into?  Given that these objects are barely
resolved in the {\em HST}\,/ACS data with $r_{hl}< 0.5$\,kpc, they are
unlikely to passively evolve into the ellipticals we see in the
present-day Universe -- these are typically larger. Merging would be
required to explain the size evolution, although we note that the
stellar masses and spatial scales of our $z\approx 6$ galaxies are
similar to those of some spiral bulges today: a stellar age of
$\approx 12\,$Gyr (Wyse, Gilmore \& Franz 1997) today would imply a
formation epoch of $z>5$.

\subsection{Effects of Metallicity, Dust and IMF on SED Fitting}
\label{sec:dust}

\begin{figure}
\resizebox{0.48\textwidth}{!}{\includegraphics{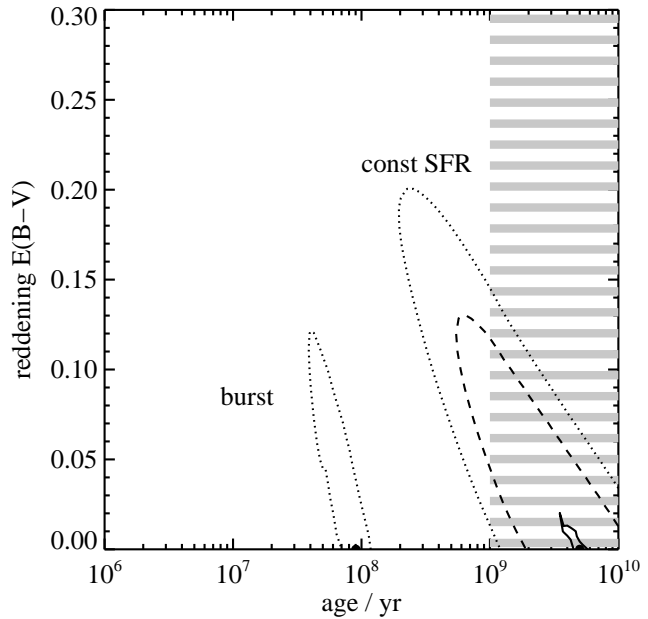}}
\caption{SBM03\#1: A plot showing the reddening E(B-V) values for the two 
limiting scenarios of an instantaneous burst model and a constant SFR model, 
for the case metallicity = $Z_{\odot}$. Contours are 68\% confidence (solid 
line), 95\% confidence (dashed line) and 99\% confidence (dotted line) for 
reduced $\chi^2$ of $\approx 1,2,3$.}
\label{fig:solardustSBM03_1}
\end{figure}

\begin{figure}
\resizebox{0.48\textwidth}{!}{\includegraphics{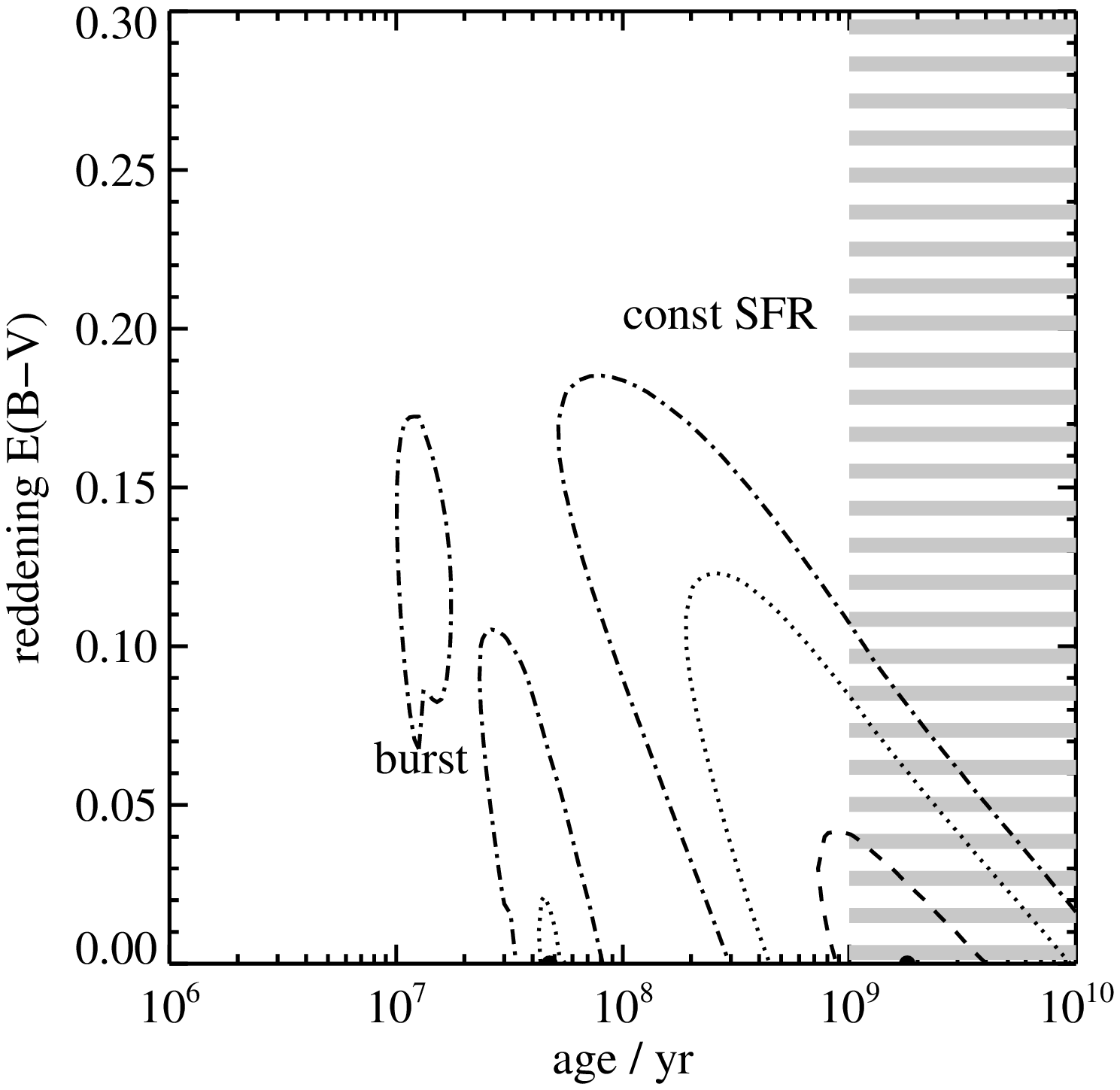}}
\caption{As Figure~\ref{fig:solardustSBM03_1}, but for SBM03\#3. Contours are for
 reduced $\chi^2$ of $\approx 1,2,3,4$, corresponding to 68\% confidence 
(solid line), 95\% confidence (dashed line), 99\% confidence (dotted line), 
and 4\,$\sigma$ confidence (dot-dash line).}
\label{fig:solardustSBM03_3}
\end{figure}

When conducting the SED fitting, two metallicities were considered:
solar ($Z_{\odot}$) and a sub-solar model ($0.2Z_{\odot}$).  The ages
and masses of the best-fit models were similar for both metallicities, 
although the sub-solar model returned slightly better fits to the data, 
with smaller reduced $\chi^{2}_{min}$ values. 

We also considered whether the red optical-infrared colours (spanning the
rest-frame UV to optical) could be attributable to dust reddening
instead of, or as well as, an underlying old stellar population. To
further examine this, we adopt the reddening model of Calzetti
(1997)\footnote{The Calzetti reddening is an empirical law is given in
  terms of the colour excess $E(B-V)=0.44\,E_{\rm gas}(B-V)$ with the
  wavelength dependence of the reddening expressed as:
\[
k(\lambda)=2.656(-2.156 + 1.509/\lambda - 0.198/\lambda^{2} +
0.011/\lambda^{3}) + 4.88
\]
for  $0.12\,\mu{\rm m}\le\lambda\le 0.63\,\mu{\rm m}$, and
\[
k(\lambda)=[(1.86 - 0.48/\lambda)/\lambda - 0.1]/\lambda +
1.73 \]
for $0.63\,\mu{\rm m}\le\lambda\le 1.0\,\mu{\rm m}$,
and the flux attenuation is given by:
\[
F_{\rm obs}(\lambda)=F_0(\lambda)\,10^{-0.4E(B-V)k(\lambda)}
\]
}, appropriate to starburst galaxies.  We constructed a grid of models
for both the SSP (instantaneous burst) and constant SFR scenarios,
using the Bruzual \& Charlot templates over the full range of 221 age
steps. For each age, we varied reddening in the range $E(B-V)= 0.00 -
1.00$\,mag, in steps of $0.01$\,mag, computing the reduced $\chi^2$ at
each step.  We find little evidence for substantial dust reddening in
the detected starlight out to $\approx 5\,\mu$m: the formal best-fits
for both SBM03\#1\,\&\,\#3 are no reddening. This is consistent with
the flat spectral slopes in $f_{\nu}$ between the ACS $z'$-band and
the NICMOS $J$- and $H$-bands reported by Stanway, McMahon \& Bunker
(2005) for many of the $i'$-drops in the {\em Hubble} Ultra Deep
Field, including SBM03\#1. The brightening in $f_{\nu}$ flux between
the near-infrared ($0.8-2.2\,\mu$m) and the IRAC bands at $3-5\,\mu$m is
best explained by a spectral break rather than the smoother continuum
gradient produced by dust reddening, and this is reflected in the
best-fit stellar populations.

Finally, we tested the effects on the derived stellar masses of the
assumed initial mass function. For the galaxy SBM03\#1 (where the
multi-waveband data was best, as it included NICMOS imaging in the
near-infrared), we re-ran the `two-population' model fits
(Section~\ref{sec:age}) with a Chabrier (2003) IMF, instead of a
Salpeter (1955) power law IMF. We used the same ongoing 10\,Myr
constant star formation rate burst, and fit for the age of an
underlying older stellar population and the relative stellar masses of
the burst and the old stars. Both IMFs produced comparably good fits
(near-identical $\chi^2$ values), the same 450\,Myr ages, and
near-identical burst fractions (0.6\% and 0.7\% by mass for the
Chabrier and Salpeter IMFs). The main difference came in the
best-fitting total stellar mass: the Chabrier model produced a mass
$\approx 30$\% less than the Salpeter IMF ($2.5\times
10^{10}\,M_{\odot}$ compared with $3.6\times 10^{10}\,M_{\odot}$ for
SBM03\#1). This effect is mainly a mass re-scaling independent of star
formation history, as the discrepancy primarily arises from the
different mass fraction in long-lived low-mass stars.

\subsection{Star Formation Rates}
\label{sec:SFRs}

We can usefully compare the ongoing star formation rate, the stellar
mass and implied age, to deduce whether our two selected $z\approx 6$
galaxies are being seen during a fairly quiet or active period in
their history.

First, let us consider the ongoing SFR. Star formation will dominate
the rest-frame UV light (probed by the {\em HST}/ACS images) in the
absence of dust obscuration or a significant AGN contribution.  In our
first analyses of $i'$-drop galaxies in the GOODS ACS images (Stanway,
Bunker \& McMahon 2003; Stanway et al.\ 2004a) and the Ultra Deep Field
(Bunker et al.\ 2004), we used the conversion from rest-frame UV flux
density to SFR, given by Madau, Pozzetti \& Dickinson (1998),
appropriate for continuous star formation. Based on the $z'$-band
magnitudes we inferred unobscured SFRs of
$19.5$\,\&\,$33.8\,M_{\odot}\,{\rm yr}^{-1}$ for SBM03\#1\,\&\,\#3,
after accounting for Lyman forest blanketting of $D_A\approx 0.95$
shortwards of Lyman-$\alpha$.

The fits of the Bruzual \& Charlot stellar synthesis models to the
broad-band photometry provide estimates of the current SFR for a range
of histories
(Tables\,\ref{tab:SBM03_1_solar},\,\ref{tab:SBM03_1_subsolar},\,\ref{tab:SBM03_3_solar},\,\ref{tab:SBM03_3_subsolar}\,\&\,\ref{tab:tau}).
However, a firm lower limit arises from our Keck/DEIMOS spectroscopy
(Bunker et al.\ 2003, Stanway et al.\ 2004a), which revealed
Lyman-$\alpha$ emission. If this line emission is powered by ionizing
photons from OB stars, then there must be star formation activity in
these galaxies within the past 10\,Myr (the lifetime of these stars,
which dominate the Lyman continuum flux).

We have argued previously that the absence of X-ray emission
and high-ionization lines such as N{\scriptsize V}\,1240\,\AA\,,
coupled with the narrow velocity width of the Lyman-$\alpha$ emission,
renders an AGN interpretation unlikely. Our spectroscopy of
SBM03\#1\,\&\,\#3 indicates line fluxes of $2\times 10^{-17}\,{\rm
  erg\,s^{-1}\,cm^{-2}}$ for both sources, with rest-frame equivalent
widths of $W_{\rm rest}=30,20$\,\AA.  Assuming case~B recombination
and the same Salpeter IMF as in the Bruzual \& Charlot models, the
star formation rates from Lyman-$\alpha$ are $\approx
6\,M_{\odot}\,{\rm yr}^{-1}$. This may be treated as a firm {\em lower
  limit} on the current SFR, as the resonantly-scattered
Lyman-$\alpha$ line is invariably quenched by dust to well below its
case~B line strength.
  
Our spectroscopic lower limit of an ${\rm SFR}>6\,M_{\odot}\,{\rm
  yr}^{-1}$ rules out the simple SSP model and those declining star
formation rate models with the shortest decay times ($\tau
=10-30$\,Myr, Figure~\ref{fig:massplotSBM03_1}). Our favoured models
are an exponentially decaying SFR with decay time $\tau\approx
300-500$\,Myr, and two-component models with $1-2$\% of the stellar
mass created in an ongoing starburst which began $10-100$\,Myr prior 
to the epoch of observation. These all
indicate similar SFRs to our original estimates of $\approx
20-30\,M_{\odot}\,{\rm yr}^{-1}$ from the $z'$-band flux.
  
Returning to the ratio of the current SFR to the stellar mass already
formed, it is helpful to consider the `$b$-parameter' (e.g.,
Brinchmann et al.\ 2004) -- a measure of the fraction of the total
stellar mass currently being born as stars, defined as $b={\rm
  SFR}/M_{\rm stellar}$. We measure $b\approx 5\,\&\,10\times
10^{-10}\,{\rm yr}^{-1}$ for SBM03\#1\,\&\,\#3, comparable with that
inferred by Egami et al.\ (2005) from the {\em Spitzer} image of a lensed
galaxy with a photometric redshift of $z\approx 7$ (Kneib et al.\ 
2004), and an indication of vigorous current star formation which
supports the conclusion of this paper.  However, the past-average SFR
must actually be {\em greater} than the current SFR in order to build
our $\approx 2\times 10^{10}\,M_{\odot}$ galaxies, given the short
time available prior to $z\approx 6$ (1\,Gyr). This is why decaying
SFR models appear to give the best fits to the SEDs.  Juneau et al.\ 
(2005) define a characteristic growth timescale (in a study of $z\sim
1$ galaxies), with $t_{\rm SFR}=M_{\rm stellar}/{\rm SFR} = 1/b$. The
ratio of this to the Hubble time at that epoch, $t_H(z)$, can be
interpreted as galaxies in a declining or quiescent star formation
mode if $t_{SFR}>t_H$.  This is the case for both SBM03\#1\,\&\,3 for
the best-fitting models. Indeed, to form the $2-4\times
10^{10}\,M_{\odot}$ of stars requires an average star formation rate
of $50-100\,M_{\odot}\,{\rm yr}^{-1}$ over the redshift range
$z_f=7.5-13.5$ favoured for the formation of the old stellar component
(Section~\ref{sec:age}). Hence, $\langle {\rm SFR}_{z=7.5-13.5}\rangle
/ \langle {\rm SFR}_{z=6}\rangle \approx 2-5$; the past-average star
formation rates of SBM03\#1\,\&\,3 are factors of $2-5$ greater than
their current rates at the epoch of observation ($z\approx 5.8$).

Although our chosen galaxies are the brightest confirmed $i'$-drops in the
GOODS-South field and possibly unrepresentative, the present work does 
imply that in these galaxies at least there was
a yet earlier vigorous
phase of activity, possibly at $z>10$, which may have played
a key role in reionizing the Universe.
This may be consistent with the measurement of temperature-polarization 
correlation 
of the cosmic microwave background from the {\em Wilkinson MAP} 
satellite by Kogut et al.\ (2003).

\section{Conclusions}
\label{sec:conclusions}

Our group previously identified and spectroscopically-confirmed
$z\approx 6$ galaxies through {\em HST}/ACS $i'$-drop imaging and
Keck/DEIMOS \& Gemini/GMOS spectroscopy. This paper presents the first
infrared detections of this population using {\em Spitzer}/IRAC.  We have
significant ($\approx 10\,\sigma$, $AB\approx 24$\,mag) detections at
3.6\,$\mu$m of the Stanway, Bunker \& McMahon (2003) galaxies \#1 and
\#3 (at $z_{\rm spec}=5.83,5.78$), and a more marginal detection of
the $z_{\rm spec}=5.79$ galaxy GLARE\#3001 (Stanway et al.\ 2004b). We
also detect SBM03\#1 at 4.5\,$\mu$m (SBM03\#3 is outside the field of
view of this $4.5\,\mu$m filter).
  
We infer from Lyman-$\alpha$ emission in our discovery spectra that
there is ongoing star formation of $> 6\,M_{\odot}\,{\rm
yr}^{-1}$ (as would be expected in these rest-UV-selected objects).
However, the preceding star formation history has not been explored
until now.  In the two best-detected galaxies, we have evidence of a
significant Balmer/4000\,\AA\ break, indicative of a prominent older stellar
population which probably dominates the stellar mass. Exploring a range
of population synthesis models indicates that the average stellar age
is $>100$\,Myr; our best-fit models suggest preferred ages of
$250-650$\,Myr for an exponentially-declining star formation rate (of
decay time $\tau\approx 70-500$\,Myr) or a two-component model (with
an ongoing starburst responsible for $0.5-5$\% of the total stellar
mass). This implies formation epochs of $z_{f}\approx 7.5-13.5$ for
the galaxies SBM03\#1\,\&\,\#3.

In all our models, the best-fit stellar masses are
$>10^{10}\,M_{\odot}$, with 95\% confidence masses of
$1.3-3.8\,\times\,10^{10}\,M_{\odot}$. This indicates that at least some
galaxies with stellar masses $>20$\% the mass of $L^*$ galaxies today
were already assembled within the first Gyr of the Universe. For these
objects, the past average star formation rate is comparable to, or greater 
than the current SFR, implying that there may have been even more vigorous
episodes of star formation at higher redshifts.  These may have played
a key role in reionizing the Universe, consistent with the earlier
studies of Bunker et al.\ (2004) and Egami et al.\ (2005).

\subsection*{Acknowledgments}

We thank Karl Glazebrook, Rychard Bouwens, Richard McMahon, Rodger
Thompson and the anonymous referee for very useful comments. This
work is based [in part] on observations made with the {\em Spitzer
  Space Telescope}, which is operated by the Jet Propulsion
Laboratory, California Institute of Technology under NASA contract
1407.  Observations have been carried out using the Very Large
Telescope at the ESO Paranal Observatory under Program ID:
LP168.A-0485.  This paper is based in part on observations made with
the NASA/ESA Hubble Space Telescope, obtained from the Data Archive at
the Space Telescope Science Institute, which is operated by the
Association of Universities for Research in Astronomy, Inc., under
NASA contract NAS 5-26555. These observations are associated with
proposals \#9425\,\&\,9583 (the GOODS public imaging survey). We are
grateful to the GOODS team for making their reduced images public -- a
very useful resource.  LPE acknowledges a Particle Physics and
Astronomy Research Council (PPARC) studentship supporting this study.

\bsp

\end{document}